\documentclass[referee,useAMS,usenatbib]{mn2e}
\usepackage{graphicx}
\usepackage{hyperref}
\usepackage{epstopdf}
\usepackage{enumerate}
\usepackage{bm}
\usepackage{lscape}

\title[Very low mass star J2343$+$29B]
{Detection of a very low mass star in an Eclipsing Binary system}

\author[Chaturvedi et al.]
  {Priyanka Chaturvedi$^{1}$,
   Abhijit Chakraborty$^{1}$,
   B.G. Anandarao$^{1}$,
\newauthor
   Arpita Roy $^{2,3}$
   and Suvrath Mahadevan$^{2,3}$ 
\\
 $^1$ Astronomy \& Astrophysics Division, Physical Research Laboratory, Ahmedabad 380009, India\\ 
 $^2$ Dept. of Astronomy \& Astrophysics, Pennsylvania State University, University Park, PA 16802\\
 $^3$ Center for Exoplanets and Habitable Worlds, The Pennsylvania State University, University Park, PA 16802\\}

\date{Received 2015 XXXX XXX }

\pagerange{\pageref{firstpage}--\pageref{lastpage}} \pubyear{2015}

\def\LaTeX{L\kern-.36em\raise.3ex\hbox{a}\kern-.15em
    T\kern-.1667em\lower.7ex\hbox{E}\kern-.125emX}

\begin{document}
\label{firstpage}
\maketitle

\begin{abstract}

We report the detection of a very low mass star (VLMS) companion to the primary star 1SWASPJ234318.41$+$295556.5A (J2343$+$29A), using radial velocity (RV) measurements from the PARAS (PRL Advanced Radial-velocity Abu-sky Search) high resolution echelle spectrograph. The periodicity of the single-lined eclipsing binary (SB$_1$) system, as determined from 20 sets of RV observations from PARAS and 6 supporting sets of observations from SOPHIE data, is found to be $16.953$~d as against the $4.24$~d period reported from SuperWasp photometry. It is likely that inadequate phase coverage of the transit with SuperWasp photometry led to the incorrect determination of the period for this system. We derive the spectral properties of the primary star from the observed stellar spectra: $T_{\rm{eff}}$~=~5125$~\pm~67$~K, $[Fe/H]$~=~$0.1~\pm~0.14$ and ${\rm log}~g$~=~$4.6~\pm~0.14$, indicating a K1V primary. Applying the Torres relation to the derived stellar parameters, we estimate a primary mass $0.864_{-0.098}^{+0.097}$~M$_\odot$ and a radius of $0.854_{-0.060}^{+0.050}$~R$_\odot$. We combine RV data with SuperWASP photometry to estimate the mass of the secondary, $M_B = 0.098 \pm 0.007 M_{\sun}$, and its radius, $R_B = 0.127 \pm 0.007~R_{\sun}$, with an accuracy of $\sim$7$\%$. Although the observed radius is found to be consistent with the Baraffe's theoretical models, the uncertainties on the mass and radius of the secondary reported here are model dependent and should be used with discretion. Here, we establish this system as a potential benchmark for the study of VLMS objects, worthy of both photometric follow-up and the investment of time on high-resolution spectrographs paired with large-aperture telescopes. 
\end{abstract}


\section{Introduction}

M dwarfs make up most of the galaxy's stellar budget. However, due to their small masses and fainter magnitudes in the visible band, such systems have remained largely unexplored. The Mass-Radius (M-R) relation serves as a crucial test for the verification of theoretical models of M dwarfs against values derived from observation. This is also integral to the understanding of stellar structure and evolution. Compared to M dwarfs, stars with masses $\ge 0.6 M_{\sun}$ have a well-established M-R relation \citep{Torres2010}. The vast majority of observations of M dwarfs of varying masses have reported radii higher than those predicted by models \citep{Torres2010, Lopez-Morales2007}. Improper assumptions of opacity in M dwarf models is speculated to be one of the reasons for this mismatch between observational and theoretically predicted radii  of M dwarfs, at the level of $\sim10-15\%$. \cite{Chabrier2007} argue that the discrepancy is due to the high rotation rate in M dwarfs, or the effect of magnetic-field-induced reduction of the efficiency of large-scale thermal convection in their interior.

\begin{figure}
\centering
\includegraphics[width=5.0in]{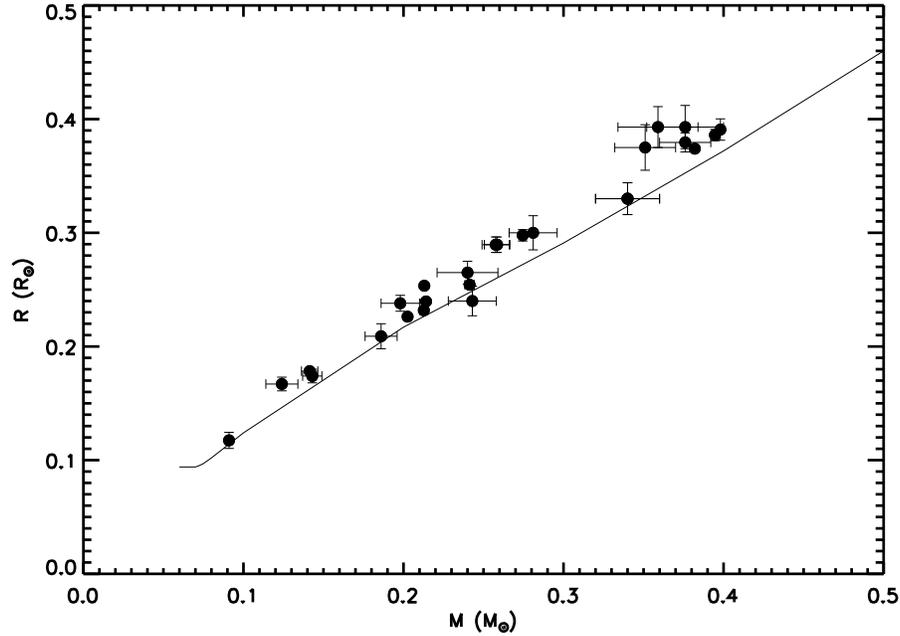}
\caption[Theoretical Mass-Radius diagram overplotted with the previously studied M dwarfs in literature and the three M dwarfs studied with PARAS]{Mass-Radius diagram for M dwarfs based on Baraffe models \citep{Baraffe2015} for 1 Gyr isochrone and solar metallicity. Overplotted are the previously studied M dwarfs listed in Table~\ref{vlms} with black filled circles with their reported uncertainties.}
\label{plot_thesis_1}
\end{figure}

M dwarfs having masses of less than $0.3 M_{\sun}$ seem to match the Baraffe models \citep{Baraffe2015} closely, as the stars become completely convective at this boundary \citep{Lopez-Morales2007}. If we choose to concentrate on very low mass stars (VLMS)~$\leq$~$0.1 M_{\sun}$, there is a dearth of such samples discovered with accuracies $\leq$ 1--2 $\%$. There have been only a handful of EB systems studied at such high accuracies in which one of the components is VLMS object with mass~$\leq$~0.1~$M_{\odot}$ \citep{Wisniewski2012, Triaud2013, Gomez2014, Ofir2012, Tal-Or2013, Beatty2007} where masses of the objects have been determined at high accuracies. Table~\ref{vlms} gives a non-exhaustive compilation of the stars having masses between $0.08-0.4$~M$_\odot$ for which masses and radii are determined at accuracies better than 10$\%$. The first three columns contain the name of the object, its mass and radius. The systems are SB$_1$ or SB$_2$ by nature. The classification of the EB, based on spectral types, is mentioned in the penultimate column. The literature references in which the sources are studied are cited in the last column of the table. A theoretical M-R diagram is plotted in Fig.~\ref{plot_thesis_1} for the M dwarfs having masses between $0.08-0.4$~M$_{\odot}$ based on Baraffe models \citep{Baraffe2015} for 1 Gyr isochrone and solar metallicity (most of the objects studied here have this age and metallicity). We have overplotted objects studied in literature from Table~\ref{vlms} with their respective error bars on masses and radii on the theoretical M-R diagram. 
As seen from Fig.~\ref{plot_thesis_1}, many of the stars studied in literature fall above the theoretical M-R plot clearly indicative of the M dwarf radius problem. Moreover, we also see that this discrepancy is more pronounced for objects having masses above 0.3~$M_{\odot}$. Out of 26 systems studied previously in literature, in the mass range of $0.08-0.4$~$M_{\odot}$ as shown in Table~\ref{vlms}, there have been only 6 systems having masses between $0.08-0.2$~$M_{\odot}$ studied with accuracies better than a few per~cent \citep{Beatty2007, Doyle2011, Triaud2013, Nefs2013, Fernandez2009}. There is thus a greater need for identifying more such VLMS candidates in EB systems and more specifically in the mass range of $0.08-0.4$~$M_{\odot}$ where there are only few sources studied.

The object chosen for the current study, J2343$+$29A, was first identified from SuperWasp (SW) photometry by \cite{Christian2006} and \cite{Cameron2007}. Observations taken from the SW North listed the primary star, J2343$+$29A, having a temperature of 5034 K, spectral type K3, a radius of ${R_{A}=~0.85~R_{\sun}}$ and a transit depth of 21 mmag. The transit light curve showed substantial scatter, and source was suspected to be a stellar binary system as per preliminary observations with SOPHIE \citep{Cameron2007}. Table~\ref{swasp} summarizes the basic stellar parameters listed for this source in the literature \citep{Cameron2007}. 

Here, we present the discovery and characterization of a VMLS companion to J2343$+$29A, enabled by the PRL Advanced Radial velocity Abu-sky Search spectrograph \citep[hereafter PARAS;][]{chakraborty14}. Spectra acquired using PARAS, over a time period of $\sim$~3 months and well-sampled in phase, are described in \S 2. In \S 3 we describe our radial velocity (RV) analysis, both independently and in concert with SW photometry. We have developed an {\tt{IDL}}-based tool {\tt{PARAS SPEC}} to determine stellar properties like $T_{\rm{eff}}$, surface gravity (${\rm log}~g$), and metallicity ($[Fe/H]$). This procedure, involving matching synthetic spectra with the observed spectra, and fitting the Fe I, Fe II absorption line equivalent width (EW), is discussed in \S 4. Also discussed in \S 4 are the stellar parameters of the primary star derived from the observed PARAS spectra. In \S 5, we discuss the implications of this work, and conclude in \S 6. 

\section{RV observations of J2343$+$29 with PARAS}

High-resolution spectroscopic observations of the star J2343$+$29A were taken with the fiber-fed echelle spectrograph, PARAS, at the 1.2 m telescope at Gurushikhar Observatory, Mount Abu, India. A set of 20 observations of the source acquired between November 2013 and January 2014 at a resolving power of 67000 were used for determining the orbital fit of the system. Details of the spectrograph, observational procedure, and data analysis techniques can be found in \cite{chakraborty14}. The spectra were recorded in the simultaneous reference mode with ThAr as the calibration lamp. All nights of observations were spectroscopic in nature. The magnitude of the source in the V band is~$\sim$~10.7, which is the limit of faintness of observations possible with PARAS on the 1.2 m telescope. The exposure time for observations was chosen to be 1800~s such that a signal-to-noise ratio (S/N) between~12--14~pixel$^{-1}$~at 5500~\AA~was obtained for each spectrum. Observations with S/N less than~11.5~pixel$^{-1}$ due to poor weather conditions or/and high air mass during observations (Air Mass~$\ge$~1.5), were not considered for orbital fitting. A list of epochs and observational details is shown in Table~\ref{rv}. The first two columns represent the observation time in UT and BJD respectively. The exposure time and observed RV are given in the following columns. The RV error based on photon noise and the uncertainties associated due to the cross correlation function (CCF) fitting are given in the last column (see \S 3 for details).

\section{Results}

\subsection{The Radial Velocity analysis}

The PARAS data analysis pipeline as described in \cite{chakraborty14} is custom-designed, fully automated and consists of robust {\tt{IDL}} routines, based on the REDUCE routines of \cite{Piskunov2002}. The pipeline performs the routine tasks of cosmic ray correction, dark subtraction, order tracing, and order extraction. A thorium line list is used to create a weighted mask which calculates the overall instrument drift, based on the simultaneous ThAr exposures. RVs are derived by cross-correlating the target spectra with a suitable numerical stellar template mask. The stellar mask is created from a synthetic spectrum of the star, containing the majority of deep photospheric absorption lines. See \citep{Pepe2002}; and references therein for a more detailed description of the mask cross-correlation method. In this case, we use a K-type stellar mask for cross-correlation. RV measurement errors are based on photon noise errors \citep{bouchy01} and the errors associated with the CCF fitting function. We randomly vary the signal on each pixel within $\pm\sqrt{N}$, where N is the signal on each pixel, and thereafter run the CCF for each spectra. This process is repeated 100 times for each spectra and the standard deviation of the distribution of the obtained RV values gives the 1~$\sigma$ uncertainty on the CCF fitting along with errors from photon noise on each RV point. The barycentric corrected RV values and their respective uncertainties are shown in Table~\ref{rv}. 

We combined PARAS observations with available SOPHIE archival data to gain a longer baseline for observations, from August 2006 to January 2014, time span of 7.5 years. We retrieved 6 observations of the star between August 2006 to August 2007 from the SOPHIE archival data~\footnote{http://atlas.obs-hp.fr/sophie/}. The SOPHIE observations for the source J2343$+$29A were obtained in the High Efficiency (HE) mode of SOPHIE, with a resolving power of $\sim$~40000 covering the wavelength region 3872-6943 \AA. SOPHIE archival data are processed by the standard SOPHIE pipeline and passed through weighted cross-correlation with a numerical mask \citep{Pepe2002}. As per the header information in the retrieved archival data, the S$/$N for each epoch varied between~$\sim$25--75~pixel$^{-1}$~at 5500~\AA. The errors on RV, as given in the header information, were calculated by the semi-empirical estimator $\sigma_{RV}~=~A~\times~\sqrt{FWHM}/(S/N~\times~C)$ as suggested by \cite{West2009} for SOPHIE data where 'A' is the empirically determined constant and 'C' is the contrast factor in the spectra. External systematic errors of 2~m~s$^{-1}$ (spectrograph drift uncertainty) and guiding errors of 4~m~s$^{-1}$ were added in quadrature \citep{Boisse2010} to the obtained statistical errors. The RV values from SOPHIE archival data and the errors on each point are given in Table \ref{sophie}. 

\begin{figure}
\centering
\includegraphics[width=5.0in]{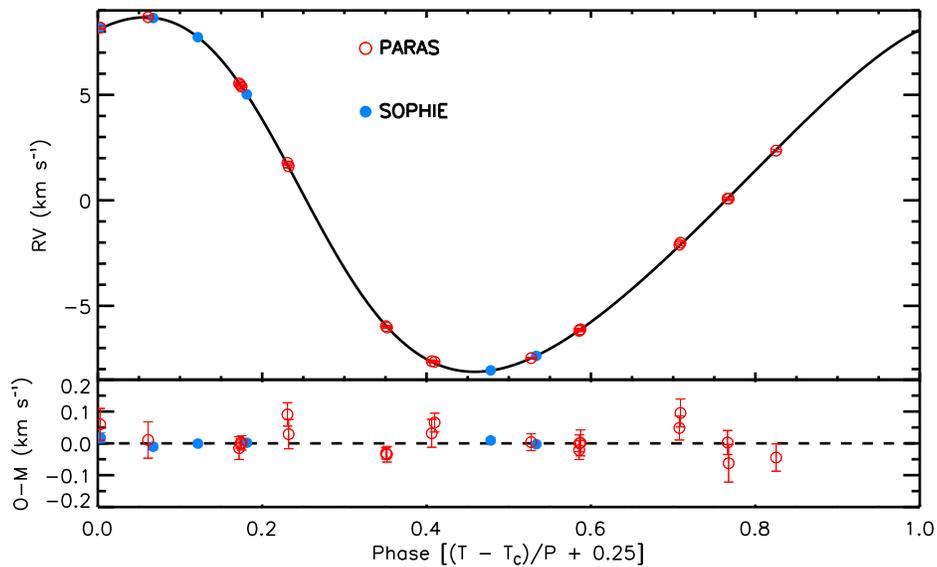}
\caption{(Top panel) PARAS, Mount Abu (open red circles) and SOPHIE (filled blue circles) observed data points along with the estimated errors are plotted. RV model curve for star 1SWASP J2334318+295556 obtained from {\tt{EXOFAST}} is overplotted against orbital phase on the observed data points. The RV model here describes the case of {\it{"RV  fit  alone"}} as described in Table 4.
(Bottom panel) The residuals from best-fitting are plotted below the RV plot. 
For better visual representation, the x axis in Phase is shifted by 0.25 so that the central primary transit crossing point (T$_c$) occurs at phase 0.25 instead of 0. P indicates the period.}
\label{rv_plot}
\end{figure}

Using {\tt{EXOFAST}} \citep{eastman2013}, we fit the spectroscopy data from the PARAS and SOPHIE spectrographs. {\tt{EXOFAST}} is a set of {\tt{IDL}} routines designed to fit transit and RV variations simultaneously or separately, and characterize the parameter uncertainties and covariances using the Differential Evolution Markov Chain Monte Carlo method \citep{johnson2011}. It requires priors on $T_{\rm{eff}}$, ${\rm log}~g$ and iron abundance ($[Fe/H]$) as an input for estimating the stellar parameters of the primary star. We provide priors of $T_{\rm{eff}}$~=~5125$~\pm~67$~K, $[Fe/H]$~=~$0.1~\pm~0.14$, and ${\rm log}~g$~=~$4.6~\pm~0.1$ to {\tt{EXOFAST}}. These priors are estimated on the basis of detailed spectroscopic analysis performed using the newly developed {\tt{PARAS SPEC}} package, as discussed in~\S 4. The RV fitting package {\tt{EXOFAST}} uses empirical polynomial relations between masses and radii of stars (for masses $\ge 0.6 M_{\sun}$); their ${\rm log}~g$, $T_{\rm{eff}}$, and $[Fe/H]$ based on a large sample of non-interacting binary stars, in which all of these parameters were well-measured \citep{Torres2010}. These priors are used as a convenient way of modelling isochrones and are fast enough to incorporate them at each step in the Markov chain. The errors are determined by evaluating the posterior probability density, based on the range of a given parameter that encompasses some set fraction of the probability density for the given model. 
It may be noted here that the RV semi amplitude (K) value as measured with {\tt{EXOFAST}} is independent of the modelled stellar parameters and has no bias contingency on any of the masses of primary or secondary stars. K value is further used independent of {\tt{EXOFAST}} to derive the mass function $f(m)$ for the EB system which is given by \citep{Kallrath2000}
\begin{equation}
 f(m)=\frac{m_{1}~\sin^3 i}{{(1+q)}^2} 
\end{equation}
Here, $m{_1}$ and $m{_2}$ are masses of the two components and q is the mass ratio of the components of EB. A knowledge of the primary mass ($m_{1}$) and $\sin i$ is needed to calculate the secondary mass ($m_{2}$). The former quantity can be derived by priors such as ${\rm log}~g$, $T_{\rm{eff}}$, and $[Fe/H]$, which are estimated by high-resolution spectroscopy as discussed in \S~4 while the latter one can be derived by photometry as presented in \S~3.2. In the singular application of RV fitting, {\tt{EXOFAST}} is equally applicable to EBs and star-planet systems. However, {\tt{EXOFAST}} uses planet approximation that ignores the mass of the secondary while calculating the orbital separation. This assumption may affect the accuracy of the derived radius of the secondary. RV values were corrected for the offset between the two spectrographs, which amounted to 259~m~s$^{-1}$. The results of this procedure are summarized in Table~\ref{result}, in the 3$^{rd}$ column \textit{``RV fit only"}. We determine the period of the EB as $16.953$~d with a RV semi-amplitude of~$8398$~m~s$^{-1}$. Fig.~\ref{rv_plot} illustrates a plot of RV versus orbital phase for J2343$+$29A. Red open circles (top panel) show RV measurements of the primary taken with PARAS, and solid blue circles denote the same for SOPHIE. The solid black curve indicates the best fit model from {\tt{EXOFAST}} based on the parameters in Table~\ref{result}. The bottom panel in Fig.~\ref{rv_plot} shows the model fit and the residuals (Observed-Model).

We tried fitting the RV data on exactly 4 times the photometry period (16.96~d) but could not constrain the fit. Thus, it should be noted that the observed period is approximately~4 times the reported period of 4.24~d from SW photometry. Since there are sufficient number of RV points, well sampled in phase of the eclipsing binary (EB), we consider the revised period of the EB from RV measurements alone to be accurate.

\subsection{SuperWASP photometry}

The light curve for J2343$+$29, as seen in Fig. 17 of \cite{Christian2006}, shows a moderate amount of scatter and an ellipsoidal amplitude of 2.9, hinting at the possibility of stellar binary \citep{Christian2006}. They identified a period of 4.24~d whereas RV data (this work) shows a periodicity of 16.953~d. The misidentification of the period by SW photometry could be due to the fact that the automated SW pipeline is intended to search for short period planets typically less than 5~d. The probability of finding transits in SW photometry, with periods that are integral multiples of 1~d or 1.5~d, is comparatively low, at around~$\sim$~35$\%$ \citep{Street2007}. Since J2343$+$29 has an RV determined period of 16.953~d (which is much higher than 5~d and also a close integral multiple of 1~d), there is greater likelihood of sampling the source poorly in phase during the entire transit duration. There are 6 transits detected for this source at $\sim$~4.24~d periodicity as reported in \cite{Christian2006}. Due to the relatively low number of recorded transits, the phase coverage of the transit event is expected to be poor.


\subsection{Simultaneous spectroscopy and photometry fitting}

\begin{figure}
\centering
\includegraphics[width=5.0in]{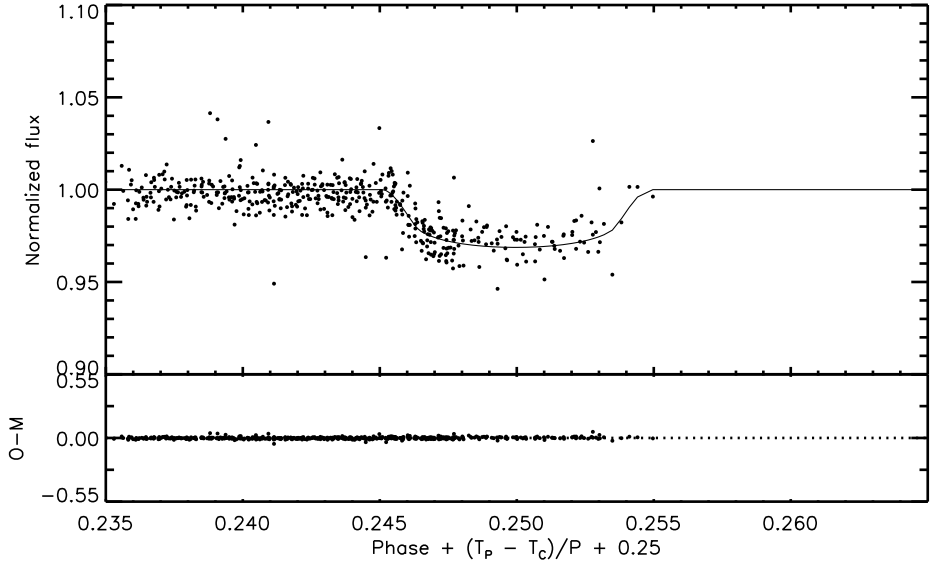}
\caption{(Top panel) Observed transit data obtained from \textit{SuperWasp} data for the period between October 2004 and December 2006. The transit model obtained from {\tt{EXOFAST}} is overplotted on the data points. The fit here shows the case of {\it{"combined  RV-Transit  photometry"}} as described in Table 4.
(Bottom panel) Observed-Fit residuals are plotted.
For better visual representation, the x axis in Phase is shifted by 0.25 so that the central primary transit crossing point (T$_c$) occurs at phase 0.25 instead of 0. P indicates the period.}
\label{transit}
\end{figure}

We fit the RV datasets and light curves simultaneously in order to impose better constraints on the execution of {\tt{EXOFAST}}. The results of the execution are summarized in Table~\ref{result} in the 4$^{th}$ column {\it{``Combined RV-transit fit"}}. We determine the period of the EB to be $16.95350\pm0.00005$~d with a RV semi-amplitude of $8407_{-10}^{+11}$ km s$^{-1}$. The orbital separation between the EB components is $0.1271_{-0.0049}^{+0.0066}$ AU. Fig.~\ref{transit} (upper panel) shows the simultaneous fit for the transit light curve obtained by analysing SW photometry data (filled circles) overplotted with the model derived from {\tt{EXOFAST}} (solid curve). The RV plot produced by simultaneous fit of RV and transit data has a close resemblance with Fig.~\ref{rv_plot}. The residuals are plotted in lower panel. The simultaneous fit gives us a transit depth of $0.025\pm0.001$~mag, angle of inclination of $89.55_{-0.40}^{+0.12}$ and a transit duration of $235\pm7$~m. Although we see a clear transit dip in the light curve, we notice that there are fewer data points close to the egress of the transit.

\section{Spectral analysis}

The double-lined EBs (SB$_2$) enable the most accurate measurements of the radii and masses of the stars. Since in such systems spectra of both the stars can be recorded, RV measurements lead to precise determination of masses of the stars in the system directly. However, in single-lined EBs (SB$_1$) systems, where the spectra of only the primary source is available, mass of the secondary is deduced by quantifying the amplitude of the wobble of the primary star in the binary system. In order to infer mass of the primary and to classify whether the star is in main-sequence or in giant or sub-giant phase, it is essential to know the surface gravity (g = GM/R$^2$), often quoted in the form of ${\rm log}~g$, surface temperature ($T_{\rm{eff}}$), and metallicity ($[Fe/H]$) of the star. A detailed study by \cite{Torres2010} on a large sample of EBs has led to the establishment of an empirical relation for the mass and radius of the stars above 0.6~M$_{\odot}$ based on the stellar parameters, i.e, ${\rm log}~g$, the $T_{\rm{eff}}$ and $[Fe/H]$. In our case, since we have high-resolution spectroscopy data obtained for RV measurements, we can use the same data to derive the stellar parameters. There are many host stars like J2343$+$29, for which spectral properties have not been studied. The quantification of these parameters is of utmost importance to draw inferences about their masses and radii. In this context, we have developed a pipeline {\tt{PARAS SPEC}}, to estimate the stellar atmospheric parameters from the analysis of stellar spectra. With high-resolution spectroscopy, it is possible to determine $T_{\rm{eff}}$ and $[Fe/H]$ (based on Fe~I and Fe~II lines) as well as ${\rm log}~g$ (based on Mg~I lines) at high accuracies. The pipeline, a set of {\tt{IDL}}-based tools, is developed to facilitate the determination of stellar properties. The determination of stellar properties from observed PARAS spectra is based on two methods. The first one involves of matching of observed spectra with synthetic spectra. The second method is based on the measurement of EW from a set of Fe I and Fe II lines in the observed spectra \citep{Blanco-Cuaresma2014}. We have developed an {\tt{IDL}}-based tool, {\tt{PARAS SPEC}}, to facilitate the process of estimation of stellar parameters from observed PARAS spectra. The tool is easy to use, and has the ability to include a variety of stellar atmosphere models, abundances or line lists, depending on the spectral type of the star. We combine the data for different epochs to obtain a higher S/N spectra for this spectral analysis. The steps used by {\tt{PARAS SPEC}} are briefly described below.

\subsection{Preliminary correction of observed spectra} 

The simultaneous ThAr mode PARAS spectra obtained for RV are also used for stellar characterization. Though PARAS has a spectral coverage of 3800--9500 \AA, we utilize only the ThAr calibrated region of 3800--6800 \AA. In this wavelength range, average separation between neighbouring orders is 17 pixels \citep{chakraborty14}. Contamination occurs where argon lines are bright enough to spill over the stellar spectra in the neighbouring order. To correct for this, we subtract Dark-ThAr (first fiber is kept dark and the second fiber is illuminated by ThAr) of similar exposure time as that of the star spectra from the star-ThAr file. This greatly mitigates ThAr contamination in the stellar spectrum within the cited uncertainties. {\tt{PARAS SPEC}} requires blaze corrected and normalized stellar spectra as input data. For this purpose, a polynomial function is fitted iteratively to an accuracy of $\sim$1~$\%$ across the stellar continuum blaze profile, ignoring absorption features in the stellar spectra for each order. The observed spectrum is then divided by the polynomial function to blaze correct and normalize it at a given epoch. All observed epochs are then co-added in velocity space to improve the S/N of the stellar spectrum. All the orders are combined and stitched to a single spectra based on S/N of the spectra. The overlapping regions are combined at the mid-point while stitching the orders.
 
\subsection{Generation of synthetic library}
\label{library}

Synthetic spectra generator code {\tt{SPECTRUM}} \citep{Gray1999}~\footnote{http://www.appstate.edu/$\sim$grayro/spectrum/spectrum276/spectrum276.html} utilizes the Kurucz models \citep{Kurucz93} for stellar atmosphere parameters. {\tt{SPECTRUM}} works on the principle of local thermodynamic equilibrium  and plane parallel atmospheres. It is suitable for generation of synthetic spectra for stars from B to mid M type. It is developed with C compiler environment with a terminal mode interface to access it. We resampled the synthetic spectra at the PARAS resolving power of $67000$. We first worked with the solar spectrum observed (at a $S/N$~$\sim$150) with PARAS in order to adjust various parameters for the library of synthetic spectra, such as, microturbulence ($v_{\rm{micro}}$), macroturbulence ($v_{\rm{macro}}$), rotational velocity ($v \sin i$), $T_{\rm{eff}}$, $[Fe/H]$ and ${\rm log}~g$. When all the parameters are kept free, the best-derived model having the least $\chi^2$ for the PARAS observed solar spectra has the following values for various parameters: $v_{\rm{micro}}$~=~0.85~km~s$^{-1}$ and $v_{\rm{macro}}$~=~2~km~s$^{-1}$. The value for $v_{\rm{micro}}$ obtained here is in close agreement with the one derived by \cite{Blackwell1984}. The value of $v_{\rm{macro}}$~=~2~km~s$^{-1}$ derived here is consistent with the value of 2.18~km~s$^{-1}$ obtained by \citep{Valenti1996}. A library of $19,200$ synthetic spectra are generated based on different combinations of $T_{\rm{eff}}$, $[Fe/H]$, ${\rm log}~g$, and $v \sin i$. The generated synthetic library consists of a coarse grid that was interpolated on-the-fly while running iterations, in an effort to increase the precision of derived parameters. The library ranges in $T_{\rm{eff}}$ between $4000-7000$ K at a temperature interval of $250$ K, in $[Fe/H]$ between $-2.5-0.5$ dex with an interval of $0.5$, in ${\rm log}~g$ between $3.0-5.0$ dex at an interval of $0.5$ and $ v sin i $  between $1-40$ km s$^{-1}$ at $1$ km s$^{-1}$ interval. The wavelength ranges between $5050-6560$ \AA~at $0.01$ \AA~interval.

\subsection{Methodology for spectral analysis} 

As mentioned earlier, {\tt{EXOFAST}} requires priors such as ${\rm log}~g$, the $T_{\rm{eff}}$ and $[Fe/H]$. In order to obtain these, we analysed the spectra using two methods described below.

\subsubsection{Synthetic Spectral Fitting}

In this method, the observed spectrum is matched with a library of synthetic spectra given as mentioned in \S~\ref{library} after adjusting for the instrument resolution. The spectrum having the best match (minimum $\chi^2$ value) gives the best-fit result for the spectral properties of the star. We carefully adjusted the continua by comparing observed PARAS spectra with synthetic spectra. The wavelength region of $5050-6500~\AA$ is used for this exercise, and the RMS residual $\sum\nolimits {(O(i) - M(i))}^2$ is computed. $T_{\rm{eff}}$, $[Fe/H]$ and $v \sin i$ are simultaneously determined from the entire wavelength region whereas ${\rm log}~g$ is determined from the Mg lines within $5160-5190~\AA$, after fixing $T_{\rm{eff}}$, $[Fe/H]$ and $v \sin i$ from the previous step. This method requires a minimum S/N of $\sim80$~pixel$^{-1}$ in the entire wavelength region covered. For fainter stars, we concentrate on red end between $6000-6500~\AA $ of the CCD where S/N is mostly above 80 (for stars upto 11 magnitude in V band having sufficient number of epochs co-added). 


\subsubsection{Equivalent Width Method (EW)}

The EW method works on the principle of inducing neutral and ionized iron lines to invoke excitation equilibrium and ionization balance. Here, abundances as a function of excitation potential should have no trends. Abundances as a function of reduced EW (EW/$\lambda$) should exhibit no trends and the abundances of neutral iron (Fe I) and ionized iron (Fe II) should be balanced \citep{Blanco-Cuaresma2014}. A model which satisfies the above mentioned criteria is the best-fit model as determined by this method. We measure the EW of a set of Fe I and Fe II lines from the line list of \cite{Sousa2014}. Each line is carefully inspected visually for line blends before abundances are determined through EW measurements by using {\tt{SPECTRUM}}. After careful inspection of the EW fits as discussed above, a set of EW of the fitted lines is given as an input to the {\tt{ABUNDANCE}} subroutine of {\tt{SPECTRUM}}. The subroutine uses various stellar models which are formed as a combination of different $T\rm{_{eff}}$, $[Fe/H]$, ${\rm log}~g$ and $v_{\rm{micro}}$ as given in \S~\ref{library}. With the above mentioned inputs, {\tt{SPECTRUM}} is able to compute abundance of each Fe I or Fe II line. $[Fe/H]$ is kept fixed for EW analysis, to a value previously determined through literature or by the synthetic spectral fitting method. $[Fe/H]$ and v$_{\rm{micro}}$ are degenerate \citep{Valenti2005} and thereby both the parameters cannot be kept free simultaneously. For the EW method, we fix $[Fe/H]$ to determine v$_{\rm{micro}}$ independently. Apart from the central value of $[Fe/H]$, two more iterations are executed on $\pm \sigma$ value of $[Fe/H]$, thereby error on each parameter is determined. Thus, the set of parameters where the slopes and the difference are simultaneously minimized give us the best-determined $T_{\rm{eff}}$, v$_{\rm{micro}}$) and ${\rm log}~g$. 

\subsubsection{Uncertainties and limitations of the method}

The synthetic spectral fitting method yields reliable results only for spectra having S/N per pixel $\geqslant$~80. Between S/N $80-100$, the wavelength region of $6000-6500~\AA$~can be used for stellar property estimation. However, we lose information on ${\rm log}~g$ which is determined by Mg~I lines ($5160-5190~\AA$). For such cases, the EW method which works for spectra having S/N per pixel $\geqslant$~50 can be used to determine stellar properties. For, S/N $\geqslant$ 100, both the methods work well to determine all three stellar parameters, $T\rm{_{eff}}$, $[Fe/H]$ and ${\rm log}~g$. If the S/N per pixel at 6000~\AA~is above 120, the uncertainities in $T\rm{_{eff}}$ and ${\rm log}~g$ are $\pm25$ K and $\pm0.05$ respectively for both the methods. For S/N per pixel (at 6000~\AA~) between 80--100, the uncertainties are $\pm50$ K and $\pm0.1$ respectively. Similar numbers for S/N per pixel between 50--80 are $\pm100$ K and $\pm0.1$. 
Apart from these uncertainties, there could be systematic errors introduced for each of the methods. Detailed systematic error analysis for these methods is beyond the scope of the current paper. Thus, we have referred to the similar work done by \cite{Blanco-Cuaresma2014}. As discussed in the paper, for synthetic spectral fitting method, there would be systematic errors due to different kind of models considered, different linelists used for the generation of synthetic spectra, and the choice of consideration of elements (iron or all elements) used for the fitting. 
All these factors on average give rise to uncertainties $\sim$37 K in $T\rm{_{eff}}$, $\sim$0.07 in ${\rm log}~g$ and $\sim$0.05 in $[Fe/H]$. 
We have further observed that there are additional systematic errors for stars having S/N per pixel less than 100 due to improper stellar continuum estimation which is of the order of the grid size of the synthetic spectra library. Moreover, there could be systematic errors introduced if a smaller wavelength region (6000--6500~\AA) (for S/N $<$ 100) is used for the estimation of stellar parameters instead of the entire wavelength region (5050--6500~\AA). Such uncertainties are $\sim$50 K in $T\rm{_{eff}}$ and $\sim$0.1 dex in ${\rm log}~g$. Considering all these factors, the systematic uncertainties for stars having S/N less than 100 is around 67 K in $T\rm{_{eff}}$, 0.11 in ${\rm log}~g$ and 0.11 in $[Fe/H]$ in case of spectral fitting method. EW method could also have systematic errors due to different kinds of models and linelists used for the estimation of EWs, and due to the rejection of outliers during linear fitting for the slope determination. These systematic errors as interpreted from \citep{Blanco-Cuaresma2014} are of the order of 45 K in $T\rm{_{eff}}$ and 0.1 in ${\rm log}~g$. The stellar parameters derived along with their combined formal and systematic uncertainties for each of the methods are listed in Table~\ref{resultspectra}.

\subsection{Results obtained from {\tt{PARAS SPEC}}}

Both the synthetic spectral fitting and EW methods are applied on several known stars. The results are summarized in Table \ref{resultspectra}. The routine is executed on the wavelength region between 5050-6500~\AA~for few stars having higher S/N per pixel but the results remain the same even if the stars are executed for a shorter wavelength range. 
We applied both the methods on our target of interest, J2343$+$29. Since, relatively high S/N spectra from SOPHIE were readily available in the archive, we applied  {\tt{PARAS SPEC}} pipeline to both PARAS and SOPHIE spectra.

\subsubsection{PARAS spectra}

\begin{figure}
\centering
\hbox{
\includegraphics[width=5.0in]{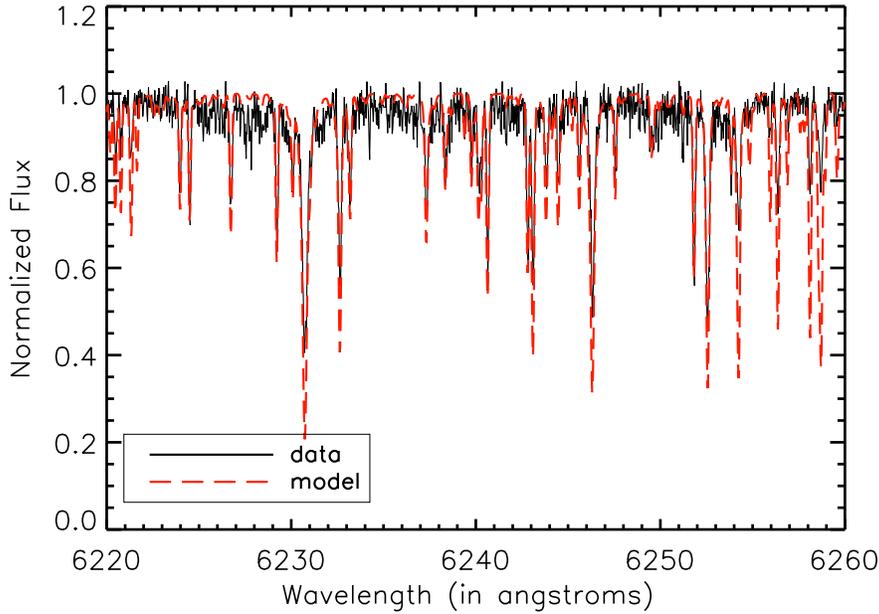}
}
\caption{Observed normalized spectra for J2343$+$29 (solid black line) plotted across the wavelength region of 6220--6260~\AA. Overplotted is the modelled spectra (red dash line) obtained from {\tt{PARAS SPEC}} analysis, with temperature value of $T\rm{_{eff}}$ of 5125 K, $[Fe/H]$ of $+0.1$ and ${\rm log}~g$ of 4.6.}
\label{swasp-paras1_spec}
\end{figure}

\begin{figure}
\centering
\includegraphics[width=5.0in]{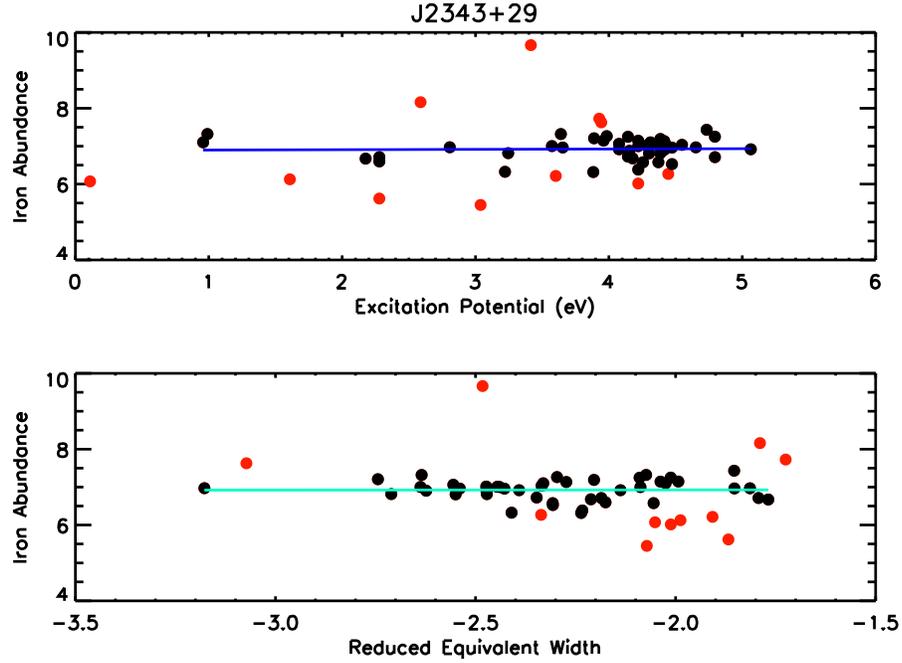}
\caption{(Top panel) Iron abundance for J2343$+$29 is plotted against excitation potential for each Fe I or Fe II line from the line list. The blue line is the fit to each data point seen in the scatter plot indicating the least slope for the best determined model temperature. (Bottom panel) Iron abundance is plotted against reduced EW and the blue-green line indicates the least slope for best determined model microturbulent velocity. The red points are the discarded points having standard deviation beyond 1 $\sigma$ (not considered for the fit). The best-fit determined parameters are: $T\rm{_{eff}}$=5125 K, $[Fe/H]$=0.1, ${\rm log}~g$ = $4.6$, $v_{\rm{micro}}$=1.2~km s$^{-1}$}
\label{ew_temp_swasp1}
\end{figure}

We co-added 28 PARAS observed spectra for the source to get a S/N of 65-70 in the blue end (5000--6000~$\AA$). The S/N, as mentioned in \S 4.3.1, is insufficient for the synthetic spectral fitting method to operate in the blue end. Hence, we concentrated on the red end of the CCD (6000--6500~$\AA $) where the S/N for the co-added spectra is~$\sim$~85~pixel$^{-1}$. We determined $T_{\rm{eff}}$ and $[Fe/H]$ by the synthetic spectral fitting method but ${\rm log}~g$ could not be determined by this method due to omission of the wavelength region having Mg lines (5160--5190~$\AA $). The stellar parameters obtained from spectral fitting are $T_{\rm{eff}}=5100\pm84$~K and $[Fe/H]=0.1\pm~0.14$. We then applied the EW method on the star J2343$+$29 and calculated $T_{\rm{eff}}$, ${\rm log}~g$ and v$_{\rm{micro}}$. The spectral properties determined by PARAS SPEC EW method for J2343$+$29 are: $T_{\rm{eff}}=5125\pm67$~K, $[Fe/H]=0.1\pm~0.14$, ${\rm log}~g=4.6\pm0.14$ and v$_{\rm{micro}}=1.2~\pm$0.1~km~s$^{-1}$. The $T_{\rm{eff}}$ determined for the star is close to the photometrically derived $T_{\rm{eff}}$ as given in \cite{Cameron2007}. The best-fit model spectra determined for J2343$+$29 is shown in Fig.~\ref{swasp-paras1_spec}. The black solid line indicates the observed normalized spectra from PARAS, and overlaid red dashed line is the best-fit model determined from this work. In Fig.~\ref{ew_temp_swasp1}, a least squares slope ($0.01_{-0.01}^{+0.02}$) for iron abundances vs. excitation potential indicative of best-fit $T_{\rm{eff}}$ for J2343$+$29 is shown in the upper panel. In the bottom panel, a plot of iron abundance vs reduced EW is shown for the least square slope ($0.005_{-0.005}^{+0.009}$) for best-fit v$_{\rm{turb}}$. 

\subsubsection{SOPHIE spectra}

We used a single SOPHIE reduced spectra having a S/N above 80 in the wavelength region 6000--6500~\AA. The synthetic spectral fitting was utilized to determine the $T_{\rm{eff}}$=$5000\pm84$~K and $[Fe/H]=0.2\pm~0.14$. We also worked with the EW method on the star J2343$+$29 and the spectral properties obtained are: $T_{\rm{eff}}$~=~5150$~\pm~67$~K, $[Fe/H]$~=~$0.2~\pm~0.14$, ${\rm log}~g$~=~$4.5~\pm~0.14$ and v$_{\rm{micro}}$=1.2~$\pm$0.1~km~s$^{-1}$. Thus the results derived for the stellar parameters by both the methods (EW and spectral fitting) from both PARAS and SOPHIE spectra are consistent within error bars as shown in Table \ref{resultspectra}. 
It is to be noted that the results obtained from EW method on the PARAS spectra have been used for further analysis. 

\section{Discussion}

Based on spectral analysis and models from {\tt{EXOFAST}}, we have determined the mass and radius, and thereby the spectral type of the primary star to be K1. With an angle of inclination determined to be $89.55_{-0.40}^{+0.12}$, and an RV semi-amplitude (K) of $8407_{-10}^{+11}$ from the combined fit to spectroscopic and photometric data, we derive the mass of the secondary star to be $0.098~\pm~0.007~M_{\sun}$. 

We use the primary star's $T_{\rm{eff}}$ and spectral type to estimate the absolute magnitude of the star as $\sim6.4$ mag \citep{cox2000}. Hence, the distance to the star is $\sim$80~pc, indicative of it being in the solar neighborhood. \cite{Meibom2015} have predicted the age of Sun like stars in solar neighbourhood based on their rotational periods. In order to get a handle on the age of the system, we must determine the rotational period of the star. Based on the width of the CCF and the formulation given in \cite{Queloz1998}, we determine a $v \sin i$ of $3.2~\pm 0.5$~km s$^{-1}$ Since the radius of the primary star is $\sim0.85~R_{\sun}$, we calculate a rotational period of $13.6 \pm 0.5$~d. The calculated rotational period of the star helps us estimate the age of the star to be between $1.5 \pm 0.5$ Gyr \citep{Meibom2015}.

From \cite{Baraffe2015} models, the theoretical radius for a~$\sim$$0.098~M_{\sun}$ star turns out to be $0.12\pm0.01~R_{\sun}$. The radius of the secondary companion, determined from observations based on transit depth, is $0.127\pm0.007~R_{\sun}$. Although the observational value of radius is dependent on the accuracy of the models used, within the error bars of the orbital fitting, we can conclude that we see almost no discrepancy associated between the observed and theoretically derived radius values. To be precise, the parameters should not be used as test of models, however they are in fact consistent with what the models predict. This substantiates the claim by \cite{Lopez-Morales2007} that observations for stars sufficiently less than $0.3~M_{\sun}$ match well within the models. However, in order to examine the M-R relation in the VLMS region, it is very important to increase the statistical sample size of such objects with high-precision measurements. 
The q ($=M_2/M_1$) of this system is $\sim0.1$, which is a rarity among short orbital period EBs (see Figure 9 of \cite{Wisniewski2012}). Due to the low mass ratio, this object seems to reside in a mass ratio-period deficit for low mass stellar binaries. Short period, low q companions are more probable to be found around F type stars, rather than G or K primaries. Over the last few years, a handful of F+M binaries have been discovered and their properties determined (e.g. \citealt{Pont 2005a}; \citealt{2005b}; \citealt{2006}; \citealt{chaturvedi14}). \cite{Bouchy2011,Bouchy2011b} suggest that for a massive companion to exist around a primary star, the total angular momentum must be above a critical value. If a primary star has a smaller spin period than the orbital period of the system (as in the case for G-type stars), the tidal interactions between the two stars will cause the secondary companion to be eventually engulfed by the primary. However, this is less likely to occur among fast rotating F-type stars, which have weaker magnetic braking, and can avoid the spin-down caused by tidal effect of the massive secondary. As shown by \cite{Zahn1977} and references therein, the orbital separation of the system will decay only if the spin period of the star is larger than the orbital period of the system. Since, the rotational period of J2343$+$29 is 13.3~d, which is smaller than the orbital period of 16.953~d, we envisage a stable orbit for the system. Out of the 26 sources indicated in Table~\ref{vlms}, only three of them (marked in bold) are G/K+M systems. EB J2343$+$29 is the fourth such system studied. We emphasize that a variety of similar EBS with a VLMS component should be studied in detail to assess our understanding of evolutionary mechanism of such systems. The orbital fitting errors reported here are dependent on the models in the SB$_1$ system. In general, M dwarfs are pronounced to have the discrepancy in observationally measured and theoretically derived radius values. However, this problem is not restricted to VLMS; solar-mass stars also fail to reproduce observations to a great extent \citep{Feiden2015} and references therein. Thus, this star serves as a benchmark VLMS, and a prime candidate for future followup as an SB$_2$, to better constrain the mass and radii of the components and thereby models of stellar structure and evolution.

\section{Conclusions}

The important conclusions of this work are as follows:
\begin{enumerate}[1.] 
\item Stellar parameters determined for the primary using spectroscopic analysis suggest that the star has $T_{\rm{eff}}$~=~5125$~\pm~67$~K, $[Fe/H]$~=~$0.1~\pm~0.14$ and ${\rm log}~g$~=~$4.6~\pm~0.14$. Hence, the primary has a mass of $0.864_{-0.098}^{+0.097}$~M$\odot$ and a radius of $0.854_{-0.060}^{+0.050}$~R$\odot$. 

\item High resolution spectroscopy taken with PARAS, combined with SOPHIE archival RV data and SW archival photometric data, yield an RV semi-amplitude for the source as $8407_{-10}^{+11}$. The secondary mass from RV measurements is $M_B~=~0.098~\pm0.007~M_{\sun}$ with an accuracy of $\sim~7$~per~cent (model dependent). Hence, we conclude that J2343$+$29 is an EB with a K1 primary and a M7 secondary \citep{Pecaut2013}. The period of the EB, based on combined RV and SW photometry measurements, is revised to be 16.953~d, compared to the previously reported value of 4.24~d. 

\item We fit the light curve data simultaneously with RV data and determine the transit depth to be $25$ mmag. Based on the transit depth, the radius of the secondary is estimated as $R_B~=0.127\pm0.007~R_{\sun}$ with an accuracy of $\sim~7$~per~cent (model dependent). The observed radius is consistent with the theoretically derived radius values from Baraffe models.

\end{enumerate}

\section*{Acknowledgments}

This work has been made possible by the PRL research grant for PC (author) and the PRL-DOS (Department of Space, Government of India) grant for PARAS. We acknowledge the help from Vaibhav Dixit and Vishal Shah for their technical support during the course of data acquisition. This work was partially supported by funding from the Center for Exoplanets and Habitable Worlds. The Center for Exoplanets and Habitable Worlds is supported by the Pennsylvania State University, the Eberly College of Science, and the Pennsylvania Space Grant Consortium. We acknowledge support from NSF grants AST 1006676, AST 1126413, AST 1310885, AST 1517592 and the NASA Astrobiology Institute (NNA09DA76A) in our pursuit of precision radial velocities. This research has made use of the ADS and CDS databases, operated at the CDS, Strasbourg, France. This paper makes use of data from the first public release of the WASP data (Butters et al. 2010) as provided by the WASP consortium and services at the NASA Exoplanet Archive, which is operated by the California Institute of Technology, under contract with the National Aeronautics and Space Administration under the Exoplanet Exploration Program. We thank the anonymous referee for his/her valuable comments, which has significantly improved the paper.


\newpage
\begin{table}
\small
\caption[Literature based compilation of known VLMS]{A compilation of known VLMS other than our work for masses and radii measured at accuracies better than or at best equal to 10$\%$.} 
\label{vlms}
\begin{tabular}{lrrrr}\\
\hline
\multicolumn{1}{l}{Object} & \multicolumn{1}{r}{Mass} & \multicolumn{1}{r}{Radius} & \multicolumn{1}{r}{EB System} &\multicolumn{1}{r}{References} \\
\hline
\textbf{J1219-39B} & $0.091\pm0.002$ & $0.1174_{-0.0050}^{+0.0071}$ & \textbf{K+M} & (1) \\
HAT-TR-205 & $0.124\pm0.01$ & $0.167\pm0.006$ & F+M & (2) \\
KIC 1571511B & $0.14136^{+0.0051}_{-0.0042}$ & $0.17831_{-0.0016}^{+0.0013}$ & F+M & (3) \\
WTS19g4-020B & $0.143\pm0.006$ & $0.174\pm0.006$ & M+M & (4)  \\
\textbf{J0113+31B} & $0.186\pm0.010$ & $0.209\pm0.011$ & \textbf{G+M} & (5)  \\
T-Lyr1-01622B & $0.198\pm0.012$ & $0.238\pm0.007$ & M+M & (6) \\
\textbf{KEPLER16B} & $0.20255_{-0.00065}^{+0.00066}$ & $0.22623_{-0.00053}^{0.00059}$ & \textbf{K+M} & (7)  \\
KOI-126C & $0.2127\pm0.0026$ & $0.2318\pm0.0013$ & M+M & (8)  \\
CM Dra A & $0.2130\pm0.0009$ & $0.2534\pm0.0019$ & M+M & (9) \\
CM Dra B & $0.2141\pm0.0010$ & $0.2396\pm0.0015$ & M+M & (9) \\
T-Lyr0-08070B & $0.240\pm0.019$ & $0.265\pm0.010$ & M+M & (6) \\
KOI-126B & $0.2413\pm0.003$ & $0.2543\pm0.0014$ & M+M & (8) \\
OGLE-TR-78B & $0.243\pm0.015$ & $0.240\pm0.013$ & F+M & (10) \\
1RXSJ154727A & $0.2576\pm0.0085$  & $0.2895\pm0.0068$ &  M+M & (11) \\
1RXSJ154727B & $0.2585\pm0.0080$ & $0.2895\pm0.0068$  & M+M & (11) \\
LSPMJ1112B & $0.2745\pm0.0012$ & $0.2978\pm0.005$ & M+M & (12)  \\
GJ3236B & $0.281\pm0.015$ & $0.3\pm0.015$ & M+M & (13)  \\
LP133-373A & $0.34\pm0.02$ & $0.330\pm0.014$ & M+M & (14) \\
LP133-373B & $0.34\pm0.02$ & $0.330\pm0.014$ & M+M & (14)  \\
19e-3-08413B & $0.351\pm0.019$ & $0.375\pm0.020$ & M+M & (15)  \\
OGLE-TR-6B & $0.359\pm0.025$ & $0.393\pm0.018$ & F+M & (16) \\
GJ3236A & $0.376\pm0.016$ & $0.3795 \pm0.0084$ & M+M & (13) \\
19c-3-01405B & $0.376\pm0.024$ & $0.393\pm0.019$ & M+M & (15) \\
MG1-2056316B & $0.382\pm0.001$ & $0.374\pm0.002$ & M+M & (17)  \\
LSPMJ1112A & $0.3946\pm0.0023$ & $0.3860\pm0.005$ & M+M & (12) \\
CuCnCB & $0.3980\pm0.0014$ & $0.3908\pm0.0094$ & M+M & (18)  \\
\hline \\
\end{tabular}
\vspace{0.5cm}
\scriptsize{References: (1) \cite{Triaud2013}; (2) \cite{Beatty2007}; (3) \cite{Ofir2012}; (4) \cite{Nefs2013};\\
(5) \cite{Gomez2014}; (6)\cite{Fernandez2009};  (7) \cite{Doyle2011};  \\
(8) \cite{Carter2011}; (9) \cite{Morales2009}; (10) \cite{Pont 2005a}; (11) \cite{Hartman2011}; \\
(12) \cite{Irwin2011}; (13) \cite{Irwin2009}; (14) \cite{Vaccaro2007}; (15) \cite{Birkby2012}; \\
(16) \cite{Bouchy2005}; (17) \cite{Kraus2011}; (18) \cite{Ribas2003}\\ }
\end{table} 

\newpage
\begin{table}
\centering
\caption{Data on J2343$+$29 by SW photometry as discussed in Collier Cameron (2007)} 
\label{swasp}
\begin{tabular}{lr}
\hline \\
 Parameters     &     Value      \\
 \hline \\
V magnitude & 10.7 \\
$T_{\rm{eff}}$ & 5034 K \\
Spectral Type & K3 \\
Period (P) & $4.24098$~d \\
Transit Depth & 0.021~mag \\
Transit Duration ($t/P$) & 0.030 \\
Transit Epoch & 2453245.1886~HJD \\
$R_{A}$ & $0.85 R_{\sun}$ \\
$R_{B}$ & $1.2 R_{J}$ \\
\hline \\
\end{tabular}
\end{table}



\begin{table}
\caption{Observation log for the star J2343$+$29 with Mt Abu-PARAS. (See text for details)}
\label{rv}
\begin{tabular}{lcccc}\\
\hline
\multicolumn{1}{c}{UT Date} & \multicolumn{1}{c}{T-2,400,000} & \multicolumn{1}{c}{Exp. Time} & \multicolumn{1}{c}{RV} &\multicolumn{1}{c}{$\sigma$RV} \\
     &	(BJD-TDB) & (sec.) &  (m s$^{-1}$) & (m s$^{-1}$)  \\
\hline
2013 Nov  12 & 56609.15187  & 1800	& $-$15367  & 35 \\
2013 Nov  13 & 56610.14686  & 1800	& $-$19126  & 39 \\
2013 Nov  16 & 56613.12378  & 1800 	& $-$28527  & 43 \\
2013 Nov  19 & 56616.16537  & 1800	& $-$27088  & 29 \\
2013 Dec  16 & 56643.08719  & 1800	& $-$15443  & 15 \\
2013 Dec  16 & 56643.10986  & 1800	& $-$15517  & 22 \\
2013 Dec  17 & 56644.07896  & 1800	& $-$19293  & 45 \\
2013 Dec  19 & 56646.08593  & 1800	& $-$26868  & 17 \\
2013 Dec  19 & 56646.10865  & 1800 	& $-$26928  & 24 \\
2013 Dec  20 & 56647.09027  & 1800	& $-$28554  & 29 \\
2013 Dec  22 & 56649.08018  & 1800	& $-$28379  & 26 \\
2013 Dec  23 & 56650.07829  & 1800	& $-$27054  & 26 \\
2013 Dec  23 & 56650.10118  & 1800      & $-$27016  & 40 \\
2014 Jan  11 & 56669.09172  & 1800	& $-$23005  & 32 \\
2014 Jan  11 & 56669.11451  & 1800	& $-$22908  & 43 \\
2014 Jan  12 & 56670.08510  & 1800 	& $-$20812  & 37 \\
2014 Jan  12 & 56670.10783  & 1800	& $-$20825  & 58 \\
2014 Jan  13 & 56671.08817  & 1800	& $-$18543  & 42 \\
2014 Jan  16 & 56674.09761  & 1800      & $-$12707  & 53 \\
2014 Jan  17 & 56675.08732  & 1800      & $-$12219  & 56  \\

\hline
\end{tabular} \\
\end{table}


\begin{table}
\caption{Archival data obtained for the star J2343$+$29 from SOPHIE observations.}
\label{sophie}
\begin{tabular}{lcccc}\\
\hline
\multicolumn{1}{c}{UT Date} & \multicolumn{1}{c}{T-2,400,000} & \multicolumn{1}{c}{Exp. Time} & \multicolumn{1}{c}{RV} &\multicolumn{1}{c}{$\sigma$RV} \\
     &	(BJD-TDB) & (sec.) &  (m s$^{-1}$)  &	(m~s$^{-1}$) \\
\hline
2006 Aug  31 & 53978.50771 & 900          & -12981   & 16 \\
2006 Sep  01 & 53979.59139 & 900          & -12502   & 10  \\
2006 sep  02 & 53980.51167 & 900          & -13410   & 08  \\
2006 Sep  03 & 53981.52070 & 900          & -16122   & 10  \\
2007 Aug  30 & 54342.57369 & 1500         & -29170   & 08  \\
2007 Aug  31 & 54343.51469 & 1800         & -28481   & 06 \\
\hline
\end{tabular}
\end{table}

\begin{table}
\tabcolsep=0.1cm
\small
\caption{Results obtained from {\tt{EXOFAST}} for J2343$+$29.} 
\label{result}
\begin{tabular}{lrrrr}\\
\hline
\multicolumn{1}{c}{~~~Parameter} & \multicolumn{1}{c}{Units} & \multicolumn{1}{r}{RV fit only}  & \multicolumn{1}{r}{combined RV-Transit fit} \\
\hline
{Component A:} & \\
                       ~~~$M_{A}$\dotfill &Mass (M$\odot$)\dotfill & $0.868_{-0.05}^{+0.07}$ &  $0.864_{-0.098}^{+0.097}$\\
                     ~~~$R_{A}$\dotfill &Radius (R$\odot$)\dotfill & $0.78_{-0.1}^{+0.11}$ &  $0.854_{-0.060}^{+0.050}$\\
          ~~~$\log(g_A)$\dotfill &Surface gravity (cgs)\dotfill & $4.60_{-0.10}^{+0.098}$   & $4.559\pm0.054$\\
          ~~~$T_{\rm{eff}}$\dotfill &Effective temperature (K)\dotfill & $5141_{-100}^{+70}$  & $5150_{-60}^{+90}$\\
                          ~~~$[$Fe$/$H$]$ & Iron Abundance\dotfill & $0.089_{-0.098}^{+0.093}$  & $0.07_{-0.17}^{+0.008}$\\
{Component B:} & \\
                           ~~~$e$\dotfill &Eccentricity\dotfill & $0.1599\pm0.0011$ & $0.16100_{-0.0027}^{+0.0015}$\\
~~~$\omega_*$\dotfill &Argument of periastron (degrees)\dotfill & $78.05_{-0.58}^{+0.59}$   & $77.48_{-1.1}^{+0.85}$\\
                          ~~~$P$\dotfill &Period (days)\dotfill & $16.95347\pm0.00005$  & $16.95350\pm0.00005$\\
                   ~~~$a$\dotfill &Semi-major axis (AU)\dotfill & $0.124\pm0.002$   $0.1271_{-0.0049}^{+0.0066}$ \\
~~~$M_{B}$\dotfill &Mass (M$\odot$)\dotfill & $0.098_{-0.003}^{+0.005}$  & $0.098~\pm0.007$\\
~~~$R_{B}$\dotfill &Radius (R$\odot$)\dotfill       &               -                  & $0.127\pm0.007$\\
~~~$f(m)$ \dotfill & Mass function {$^{\star}$} (M$\sun$) \dotfill & $0.0789 \pm 0.009$  \\
      
{RV Parameters:} & \\
                         
       ~~~$T_{P}$\dotfill &Time of periastron (BJD)\dotfill & $2455033.471\pm0.003$  & $2453592.333_{-0.030}^{+0.023}$\\
                ~~~$K$\dotfill &RV semi-amplitude (m/s)\dotfill & $8399_{-10}^{+11}$  & $8407_{-10}^{+11}$\\
                   ~~~$M_{B}/M_{A}$\dotfill &Mass ratio\dotfill & $0.1124\pm0.0009$  & $0.1134_{-0.0060}^{+0.0056}$\\
           ~~~$\gamma$\dotfill &Systemic velocity (m/s)\dotfill & $-21017\pm12$  & $-21031.4_{-19.}^{+18.}$\\
           
{Transit Parameters:} & \\
            ~~~$T_C$\dotfill &Time of transit (BJD)\dotfill & -  & $2453592.7443_{-0.0032}^{+0.0027}$\\
~~~$R_{B}/R_{A}$\dotfill &Radius of secondary in stellar radii\dotfill & -  & $0.1471_{-0.004}^{+0.005}$\\
~~~$i$\dotfill &Inclination (degrees)\dotfill & -  & $89.55_{-0.40}^{+0.12}$\\
~~~$a/R_{A}$\dotfill &Semi-major axis in stellar radii\dotfill & -  & $32.1_{-1.7}^{+1.4}$ \\
~~~$\delta$\dotfill &Transit depth\dotfill & - & $0.025\pm0.001$\\
             ~~~$T_{14}$\dotfill &Total duration (minutes)\dotfill & -  & $235\pm7$\\
\hline
\end{tabular}\\
\end{table}





\begin{landscape}
\begin{table}
\tabcolsep=0.15cm 
\caption{Results obtained from Spectral analysis (SF=Spectral Fitting; EW=Equivalent Width Method; LV= Literature Value (Errors represented on each parameter are model dependent as discussed in text). S/N is per pixel S/N at 6000~\AA.} 
\label{resultspectra}
\begin{tabular}{|l|c|c|c|c|c|c|c|c|c|c|c|c|r|}\\
\hline
\multicolumn{1}{|l|}{Star} & \multicolumn{1}{|c|}{V$_{\rm{mag}}$} & \multicolumn{1}{|c|}{S/N} & \multicolumn{3}{|c|}{$T_{\rm{eff}}$} & \multicolumn{3}{|c|}{$[Fe/H]$} & \multicolumn{3}{|
c|}{${\rm log}~g$} & \multicolumn{1}{|r|}{$ v sin i $} & \multicolumn{1}{|r|}{Reference}\\
\hline
& & & SF & EW  & LV & SF & EW (fixed $[Fe/H]$) & LV & SF & EW  & LV & \\
\hline
Tauceti & 3.5 & 500 & $5400 \pm 44$ & $5400 \pm 47$ & $5414 \pm 10$ & $-0.50 \pm 0.07$ & $-0.5$ & $-0.5 \pm 0.01$ & $4.40 \pm 0.09$ & $4.5 \pm 0.11$ & $4.49 \pm 0.03$ & $3.0 \pm 1.0$ & (a) \\
Sigma Draconis & 4.7 & 250 &$5450 \pm 44$ & $5475 \pm 47$ & $5400 \pm 50$ &  $-0.1 \pm 0.07$ & $-0.1$  & $-0.20 \pm 0.06$ & $4.50 \pm 0.09$ & $4.50 \pm 0.11$ & $4.5 \pm 0.05$ & $3.0 \pm 1.0$ & (b) \\
Procyon & 0.37 & 550 & $6550 \pm 44$ & $6650 \pm 47$ & $6554 \pm 18$ &  $0.0 \pm 0.07$ & $0.0$  & $-0.04 \pm 0.01$ & $3.9 \pm 0.09$ & $3.9 \pm 0.11$ & $3.99 \pm 0.17$ & $5.0 \pm 1.0$ & (a) \\
HD9407 & 6.5 & 220 & $5700 \pm 44$ & $5725 \pm 51 $ & $5661 \pm30 $ & $0.0 \pm 0.11$ & $0.0$ & $0.03 \pm 0.09$ & $4.4 \pm 0.09$ & $4.35 \pm 0.11$ & $4.42 \pm 0.11$ &  $3.0 \pm 1.0$ & (c) \\
HD 166620 & 6.4 & 160 & $5200 \pm 62$ & $5025 \pm 51 $ & $4966 \pm 205 $ & $0.0 \pm 0.11$ & $0.0$ & $-0.17 \pm 0.08$ & $4.6 \pm 0.09$ & $4.35 \pm 0.14$ & $4.45 \pm 0.17$ & $4.0 \pm 1.0$ &  (c) \\
NLTT 25870 & 10.01 & 80 & $5400 \pm 84$ & $5225 \pm 67 $ & $5326 \pm 45 $ & $0.3 \pm 0.14$ & $0.3$ & $0.4 \pm 0.07$ & - & $4.6 \pm 0.14$ & $4.45 \pm 0.08$ & $3.0 \pm 1.0$ & (d) \\
HD 285507 & 10.5 & 60 & $4650 \pm 120$ & $4450 \pm 109 $ & $4542 \pm 50 $ & $0.1 \pm 0.14$ & $0.1$ & $0.13 \pm 0.05$ & - & -{$^{\star}$} & $4.67 \pm 0.14$ & $3.0 \pm 1.0$ &  (e) \\
KID 5108214 & 7.94 & 100 & $6000 \pm 84$ & $6050 \pm 67 $ & $5844 \pm 75 $ & $0.3 \pm 0.14$ & $0.3$ & $0.2 \pm 0.1$ & $4.0 \pm 0.14$ & $3.95 \pm 0.11$ & $3.80 \pm 0.01 $ & $5.0 \pm 1.0$ & (f) \\
HD 49674 & 8.10 & 90 & $5650 \pm 84$ & $5600 \pm 67 $ & $5632 \pm 31 $ & $0.2 \pm 0.11$ & $0.2$ & $0.33 \pm 0.01$ & - & $4.35 \pm 0.14$ & $4.48 \pm 0.12$ & $3.0 \pm 1.0$ & (g) \\
J2343$+$29 (PARAS) & 10.7 & 85 & $5100 \pm 84$ & $5125 \pm 67$ & $5034$  & $0.1 \pm 0.14$ & $0.1$ & --  & -- &$4.6 \pm 0.14$ & -- &   $3.0 \pm 1.0$ & (h) \\
J2343$+$29 (SOPHIE) & 10.7 & 90 & $5000 \pm 84$ & $5150 \pm 67$  & $5034$ & $0.2 \pm 0.14$ & $0.2$ & -- & -- & $4.5 \pm 0.14$ & -- &  $3.0 \pm 1.0$ & (h) \\ 
\hline
\end{tabular}
\scriptsize {References: (a) \cite{Blanco-Cuaresma2014}; (b) \cite{Soubiran2010}; (c) \cite{Palateou2015}; (d) \cite{Butler2000}; (e) \cite{McDonald2012}; \\
(f) KEPLER CFOP (https://cfop.ipac.caltech.edu); (g) \cite{Ghezzi2014}; (h) \cite{Cameron2007} \\}
{$^{\star}$} No Fe II lines were shortlisted for EW determination. \\
\end{table} 
\label{resultspectra}
\end{landscape}



\end{document}